\documentclass [aps,prx,twocolumn,superscriptaddress,floatfix]{revtex4-2}
\usepackage{xcolor}
\definecolor{forestgreen}{rgb}{0.13, 0.55, 0.13}

\usepackage{ragged2e}
\usepackage{graphicx}
\usepackage{pbox}
\usepackage{amsmath}
\usepackage{tabularx}
\usepackage{comment}
\usepackage{amsmath, amsfonts, bm, color}
\usepackage{epsfig}

\begin{document}

%\preprint{}

\title{Non-equilibrium heating path for utrafast laser-induced nucleation of skyrmion lattices.}

% author. Explanatory text should go in the  []'s, actual e-mail

% \affiliation can be followed by \email, \homepage, \thanks as well
\author{P. Olleros-Rodr\'iguez}
    \email{ pablo.olleros@imdea.org}
%\homepage []{Your web page}
%\thanks{}
    \affiliation{IMDEA Nanociencia, Campus de Cantoblanco, 28049 Madrid, Spain.}
\author{M. Strungaru}
    \email{mss555@york.ac.uk}
    \affiliation{Department of Physics, University of York, YO10 5DD, York, United Kingdom}
\author{S. Ruta}
    \affiliation{Department of Physics, University of York, YO10 5DD, York, United Kingdom}
\author{P. Gavriloaea}
    \affiliation{Department of Physics, University of York, YO10 5DD, York, United Kingdom}
\author{P. Perna}
    \email{ paolo.perna@imdea.org}
\affiliation{IMDEA Nanociencia, Campus de Cantoblanco, 28049 Madrid, Spain.}
    \author{R. W Chantrell}
\email{ roy.chantrell@york.ac.uk}
    \affiliation{Department of Physics, University of York, YO10 5DD, York, United Kingdom}
\author{O. Chubykalo-Fesenko}
    \email{ oksana@icmm.csic.es}
\affiliation{Materials Science Institute ICMM-CSIC, Campus de Cantoblanco, 28049, Madrid, Spain.}

\date{\today}

%---------------------------------------------------------------------------
% 

%               ABSTRACT
%
%---------------------------------------------------------------------------

\begin{abstract}
We explore the helicity-independent light-induced nucleation of skyrmion lattices in ferromagnetic cobalt-based trilayers with perpendicular magnetic anisotropy. Using Atomistic Spin Dynamics simulations, we show that a high temperature excitation followed by magnon drops and their non-equilibrium relaxation, accessed by an ultrafast laser excitation with specific duration and intensity, can lead to the generation of a skyrmion lattice stable at room-temperature. The nucleation window, the topological density and the skyrmion polarity can be additionally manipulated by external magnetic fields. Our results provide insight into the non-equilibrium nature of skyrmionic excitations at non-zero temperatures and pave additional routes for their use in information technologies.
\end{abstract}
\maketitle
% References should be done using the  \cite, \ref, and \label commands
%---------------------------------------------------------------------------
%
%               INTRODUCTION
% References should be done using the  \cite, \ref, and \label commands
%---------------------------------------------------------------------------
%
%               INTRODUCTION
%
%---------------------------------------------------------------------------
%
Magnetic Skyrmions (Sk) are nanometer sized topologically protected spin-textures appearing in magnetic systems that exhibit Dzyalonshinskii-Moriya interaction(DMI)\cite{dzyaloshinskii1957ie,moriya1960anisotropic}. Due to their reduced sizes and transport properties via spin-polarised currents, magnetic skyrmions are considered promising candidates as information carriers in next-generation spintronic devices \cite{fert2017magnetic,wiesendanger2016nanoscale,krause2016spintronics,Sampaio} or novel reservoir and neuromorphic computing \cite{huang2017magnetic,bourianoff2018potential}. Among several possibilities, skyrmions in thin multilayers composed of transition metals and high spin-orbit coupling materials \cite{fert2017magnetic,Sampaio,moreau2016additive} are especially interesting due to the potential usage at room temperature (RT) and long-time thermal stability. The use of skyrmions in technological applications is constrained by the necessity to nucleate them with controlled polarization. Since in thin film multilayers small Neel skyrmions in the absence of an external field are typically metastable structures, this frequently requires special protocols such as bipolar current pulse trains \cite{Woo2016}, current injection through nanocontacts \cite{Sampaio} or specially designed patterned constrictions and injectors \cite{Jiang2015,finizio2019deterministic}.

Recently, research on optically induced ultrafast magnetisation dynamics triggered by femtosecond laser pulses in ferro- and ferrimagnetic materials is gaining considerable interest given its localised and ultrafast character\cite{Kirilyuk2010}. In this scenario, magnetisation dynamics can be induced due to purely thermal magnetic excitations by ultrafast heat produced by a linearly polarised laser pulse \cite{Ostler2012}. Particularly, it has been shown that skyrmions or bubbles can be nucleated by applying laser pulses \cite{je2018creation,finazzi2013laser}. The skyrmions densities and sizes were reported to depend on the laser parameters. Since intrinsic characteristics of skyrmions only rely on the material properties, these dependencies can be explained by thermal magnetisation reversal in a large region under the laser spot stabilised by magnetostatic interactions giving rise to large magnetic bubbles \cite{je2018creation}.

\begin{figure}[ht]
    \includegraphics[trim=0cm 0 0 0.5cm,width=\columnwidth]{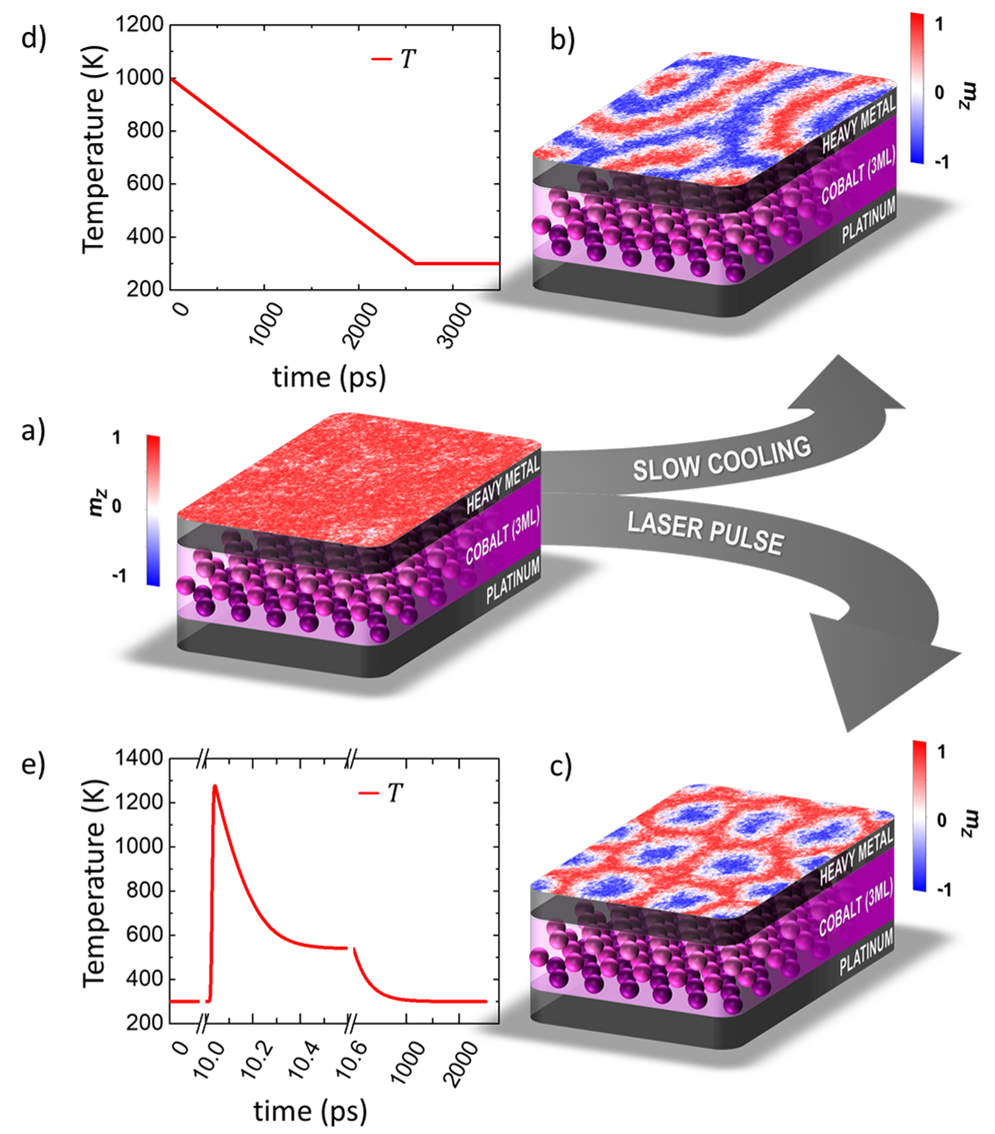}
    \caption{(Color online) {\bf Illustration of the nucleation protocols}. a) Sketch of the Pt/Co(3ML)/HM system modelled by using the parameters listed in Table 1. The top layer presents the initial spin configuration; b) Ground state obtained after slow cooling (temperature profile presented in d)) from a paramagnetic state (above system's $T_C =$ 754.29 K) to the room temperature  following a quasi-equilibrium dynamical path.  c) Skyrmion lattice obtained when a laser-pulse is applied to the system (fluence $F_0 = 5.6 mJ/cm^{2}$ and pulse-length of $10 fs$); The electron temperature profile is presented in panel e). The final temperature of both systems is $T_f=300K$. }
    \label{fig1}
\end{figure}

In the present Letter, using atomistic spin dynamics (ASD) modelling we demonstrate  the efficient generation of small skyrmions (with diameters below 20 nm) in realistically parametrised Pt/Co(3ML)/Heavy-metal(HM) magnetic trilayers \cite{moreno2016temperature,yang2015anatomy}  under the effect of non-equilibrium temperature dynamics produced by a femtosecond laser excitation. It has been shown that Co layers sandwiched between different heavy metals exhibit a strong DMI \cite{yang2015anatomy,moreau2016additive,boulle2016room} that leads to stabilisation of small skyrmions. Moreover, the large and tunable perpendicular magnetic anisotropy (PMA) of Co-based systems and the topological protection of these nanostructures can benefit their thermal stability. ASD models have been demonstrated to be a powerful technique in studying laser-induced magnetization dynamics \cite{Ostler2012,Barker2013,iacocca2019spin}, concerning especially a rapid heating to temperatures above the Curie temperature ($T_C$). In this Letter we show that a slow cooling from paramagnetic state creates a stripe-domain configuration in our modelled system. At the same time for a set of specific laser-pulse parameters it is possible to nucleate skyrmions in the absence of magnetic fields  due to the highly non-equilibrium excitation and relaxation path of magnon drops. Furthermore, we demonstrate that the skyrmion density can be manipulated by the laser properties and external magnetic fields.

The ASD simulations were performed using the software package VAMPIRE \cite{evans2014atomistic}. The spin Hamiltonian, consisting of the Heisenberg exchange, anisotropy, Zeeman and DMI energies, is written as:
\begin{equation}
\begin{split}
\label{eq1}
\mathcal{H}_{spin}=-\sum_{i<j} {J_{ij}(\mathbf{S}_{i}\cdot \mathbf{S}_{j})}-K_{u}\sum_{i}{(\mathbf{S}_{i}\cdot \hat{\mathbf{e}})^2}\quad -\\ -\mu_{S}\sum_{i}{(\mathbf{S}_{i}\cdot \mathbf{B}_{app})} -  \sum_{i<j}{\mathbf{D}_{ij} \cdot (\mathbf{S}_{i}\times \mathbf{S}_{j})}
\end{split}
\end{equation}
where ${\mathbf{S}_i}$ and ${\mathbf{S}_j}$ are unit vectors referring to the spin in the sites $i$ and $j$ respectively. $J_{ij}$ is the symmetric exchange interaction between spins $i$ and $j$, $K_u$ is the uniaxial anisotropy of the system with the easy axis pointing along the direction defined by the unit vector $\hat{\mathbf{e}}$, $\mu_S$ is the magnetic moment of the spin, $\mathbf{B}_{app}$ is the external applied field and $\mathbf{D}_{ij}$ is the DMI vector calculated as $\mathbf{D}_{ij}=D( \mathbf{z} \times \mathbf{r}_{ij})$ \cite{yang2015anatomy}, with $\mathbf{z}$ and  $\mathbf{r}_{ij}$ being the unit vectors that point along z direction and the relative distance between atoms $i$ and $j$ respectively. The modelled system consists of 3 monolayers (3ML) of hexagonal closed-packed Cobalt. The sample is parametrised using the uniaxial perpendicular anisotropy and exchange energy corresponding to bulk Co \cite{moreno2016temperature}. The DMI interaction for a Pt/Co interface is parametrised from first principles calculations \cite{yang2015anatomy} and matches that obtained experimentally \cite{ajejas2017tuning,ajejas2018unraveling} and calculated in the micromagnetic approximation \cite{olleros2020intrinsic}.
 The parameters used in the simulation are given in Table \ref{tab:tab1}.
The magnetisation dynamics is obtained by solving the set of atomistic stochastic Landau-Lifshitz-Gilbert (LLG) equations coupled to the electronic temperature $T_e(t)$. The effect of the laser pulse is included by the Two-Temperature Model (2TM)\cite{jiang2005improved} for coupled electronic $(T_e)$ and lattice $(T_l)$ temperatures dynamics  excited by a laser pulse of a power density $P(t)$:
\begin{equation}
\begin{split}
\label{eq2}
C_{e}\frac{\partial T_{e}}{\partial t} = -G_{el}(T_{e}-T_{l}) + P(t)
\\ 
C_l\frac{\partial T_{l}}{\partial t} = G_{el}(T_{e}-T_{l}) - \kappa_{e}\nabla T_{l}
\end{split}
\end{equation} 

\begin{center}
\centering
\begin{table}
\caption{\label{tab:tab1}\textbf{Magnetic and 2TM parameters used for modelling Co trilayers.}}
\centering
\fontsize{8}{6}
\begin{tabularx}{\columnwidth}{c c}
    \hline
    \toprule
    \multicolumn{2}{c}{Magnetic parameters (Co)} \\ \hline
    Exchange energy, $J_{ij}$ & $4.8\times 10^{-21} J/atom$ $^{(a)}$\\
    DMI energy, $D$ & $4.8\times 10^{-22} J/atom$ $^{(a)}$ \\
    Uniaxial Anisotropy, $K_{u}$ & $5.85\times 10^{-24} J/atom$ $^{(b)}$\\
    Damping,  $\alpha$ & $0.3$ $^{(c)}$\\
\end{tabularx} 
\begin{tabularx}{\columnwidth}{c c}
    \hline    
    \toprule
    \multicolumn{2}{c}{2TM parameters} \\
    \hline
    Electron specific heat parameter,  $\gamma_{sp}$ & $662^{ } J/m^{3}K^{2}$ $^{(d)}$\\
     Phonon specific heat, $C_{l}$ & $2.07\times 10^{6} J/(m^{3}K)$ $^{(d)}$ \\
     Electron-phonon coupling, $G_{el}$ & $4.05\times 10^{18} J/(sm^{3}K)$  $^{(d)}$\\
     Heat-sink coupling, $\kappa_{e}$ & $4\times 10^{9} s^{-1}$ $^{(e)}$
    \\\hline
    \multicolumn{2}{l}{\textsuperscript{(a)} Reference \cite{yang2015anatomy}}\\
    \multicolumn{2}{l}{\textsuperscript{(b)} Reference \cite{moreno2016temperature}}\\
    \multicolumn{2}{l}{\textsuperscript{(c)} Reference \cite{metaxas2007creep}}\\
    \multicolumn{2}{l}{\textsuperscript{(d)} Reference \cite{chimata2012microscopic}}\\
    \multicolumn{2}{l}{\textsuperscript{(e)} Reference \cite{bigot2005ultrafast}}\\
\end{tabularx}
\end{table}
\end{center}
where $C_{e}=\gamma_{sp}\cdot T_{e}$ is the electronic specific heat, $C_{l}$ is the lattice specific heat, $G_{el}$ the electron-phonon coupling and $\kappa_{e}$ the diffusion coefficient representing the heat dissipation  to the environment.
The laser power density is considered with a Gaussian form $
    P(t)=[2F_{0}/(\delta t_{p}\sqrt{\pi/ln2})]\cdot exp[(-4ln2)(\frac{t}{t_p})^2]$
where $F_{0}$ is the laser fluence (in units of energy density), $t_{p}$ the pulse temporal width and $\delta$ the optical penetration depth, assumed to be $\delta =$ 10 nm. Note that the deposited energy is independent of the pulse width for a given laser fluence. The 2TM parameters used in the simulation are given in Table \ref{tab:tab1}.

Fig.\ref{fig1} illustrates the nucleation protocols used in this work for skyrmion generation in zero external field; Panel a) shows schematically the modelled Co-based trilayer. Following a quasi-equilibrium dynamical path, i.e., a slow cooling process from temperature $T>T_C$ (See Appendix A) to RT as indicated in Fig.\ref{fig1}d, the final magnetic configuration (the ground state) consists of labyrinthine domains (Fig.\ref{fig1}b) as typically obtained in systems with PMA. On the contrary, starting from a saturated state (Fig.\ref{fig1}a) the application of a laser pulse, shown in Fig.\ref{fig1}e excites highly non-equilibrium electronic states (leading to a non-equilibrium spin dynamics) which is responsible for the nucleation of a skyrmion lattice shown in panel c). Note that although in both cases the final temperature of the systems is RT, the nucleation of skyrmions is only achieved with ultrafast temperature dynamics resulting from the application of femtosecond laser pulse, proving that skyrmion lattice is a metastable configuration and that non-equilibrium dynamical path is responsible for its generation. This is different from modelling results of Ref.\cite{koshibae2014creation,lepadatu2020emergence} where skyrmions were a ground state created due to a phase transition from ferromagnetic to the paramagnetic state and consequent cooling in which case there is no necessity for specific ultrafast non-equilibrium excitation.

To quantify the topology of the structures created by the laser pulse, we evaluate the total topological charge $Q$ of the simulated system. In the continuous approach this is given by: $Q=\frac{1}{4\pi}\int \textbf{S}(\delta_x\textbf{S} \times \delta_y\textbf{S}) dx dy$. For a discrete lattice, the topological charge is calculated as in \cite{rozsa2016complex}, based on the sum of spherical areas given by sets of 3 neighbouring spins placed in a triangle. For a skyrmion lattice with a well-defined chirality of the domain walls and a unique skyrmion polarization $P_{Core}=\pm{1}$, the total topological charge of the system characterizes  the number of nucleated skyrmions $N_{Sk}$, i.e. $Q=N_{Sk}\cdot P_{Core}$. For the numerical characterisation of the nucleated skyrmion lattices, independent on the simulated system size, we compute the topological charge surface density.

\begin{figure}[htb]
    \includegraphics [trim=0 0 0 0,width=\columnwidth]{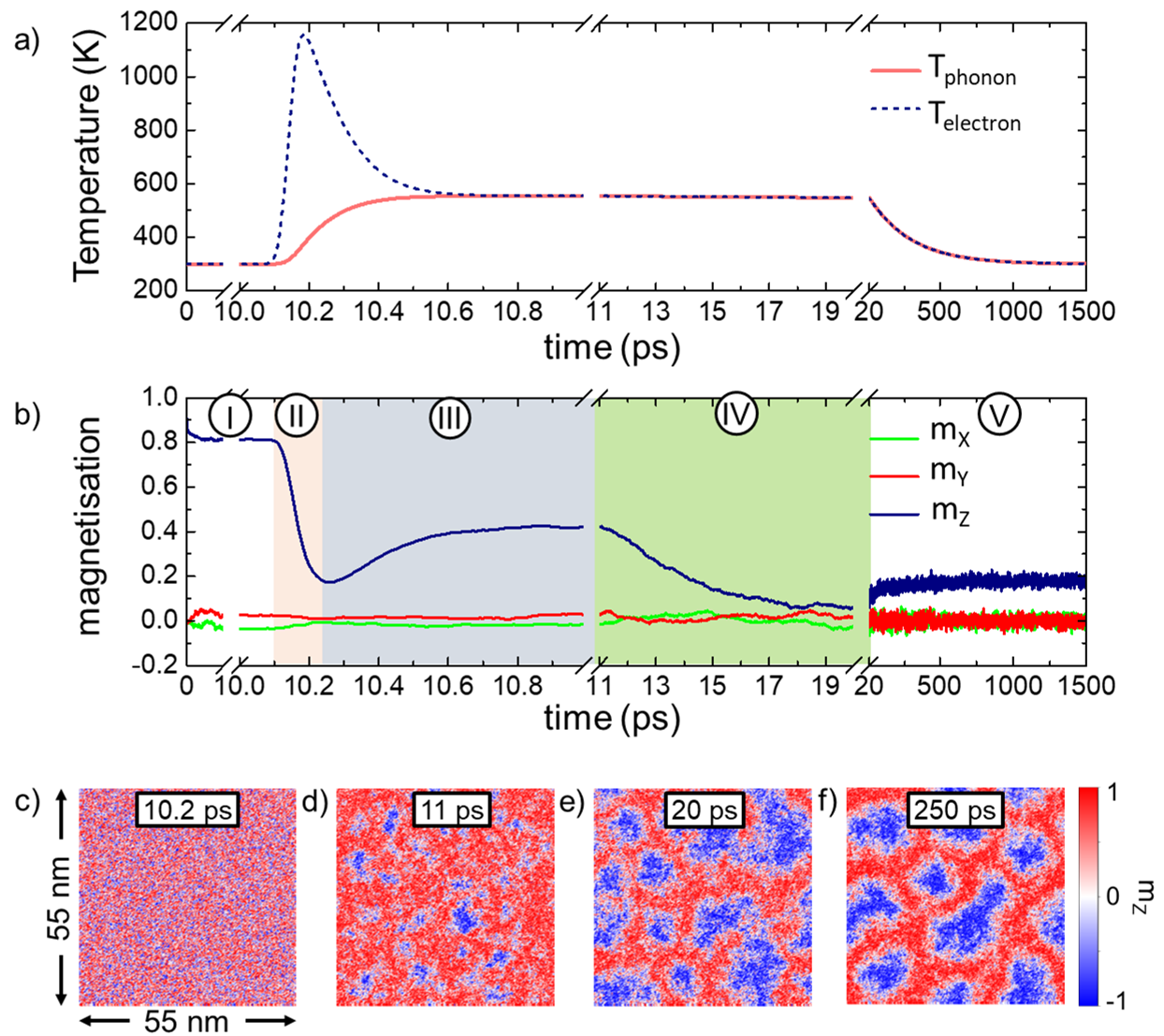}
    \caption{(Color online) {\bf Laser-induced magnetisation dynamics}. Panel a) Phonon (red solid line ) and electron (blue dashed line) temperature dynamics during the simulated light induced nucleation of skyrmion lattices (laser fluence $F_0=6$ mJ/m\textsuperscript{2} and a pulse width $t_p=50$ fs). Panel b) Dynamics of magnetisation components during the light induced nucleation of skyrmion lattice. It can be separated in five regions of interest: (I) Thermal equilibration of the initial OOP saturated state. (II) Demagnetisation process induced by the increase of the electronic (spin) subsystem. (III) Remagnetisation and nucleation of  \textit{magnon drops}. (IV) Growth and stabilisation of the magnon drops during the \textit{magnon coalescence}. (V) Room-temperature thermalisation of the system. Panels c-d) OOP-Spin configurations of the 55nm\textsuperscript{2} simulated thin film at selected times (c) $t=10.2$ps; demagnetized state (d) $t=11$ps;  the MDs localisation (e) $t=20$ ps; magnon coalescence (f)  $t=250$ps; the skyrmion lattice stabilisation.}
    \label{fig2}
\end{figure}

In our simulations we start from an initial saturated state with the magnetisation pointing  along the out-of-plane (OOP) direction (as indicated in Fig.\ref{fig1}a) and equilibrate the system at RT ($T=300$ K).  Then, we apply a non-polarised laser pulse of fluence $F_0$ and temporal width $t_p$. Fig.\ref{fig2} shows the temperature profile (Panel a) and the magnetisation dynamics (Panel b) following the application of a laser pulse of $t_p$= 50 fs and $F_0$= 6 mJ/m\textsuperscript{2}.  Panels c)-f) represent the spin configuration of the system at selected times during the process. After equilibrating the system for 10 ps (Region I, left white area in Fig.\ref{fig2}b) the laser pulse is applied, leading to an increased electronic temperature. The highly excited electrons  change the spin thermal reservoir driving the system to quasi-demagnetised state (Region II, yellow area in  Fig.\ref{fig2}b). Eventually a mutual thermalisation between the electronic, lattice and spin subsystems takes place on the timescale of hundreds of femtoseconds (Region III, grey area in Fig\ref{fig2}b). Finally, in Region IV (green area) and V(extreme right white region), the heat diffusion dissipates the deposited energy outside the system leading to its thermalisation at RT. 

Similarly to the multi-scale domain nucleation processes in ferrimagnetic alloys described in \cite{iacocca2019spin}, we can separate the ultrafast dynamics and the subsequent skyrmion lattice formation in 2 stages. First, the laser pulse rapidly quenches the magnetisation (see Fig.\ref{fig2}c). However, as was pointed out in models of ultrafast magnetisation dynamics\cite{Barker2013,kazantseva2007slow}, the spin correlations are not destroyed. During the so-called \textit{magnon localisation} process (Region III, grey in Fig.\ref{fig2}b, and Fig.\ref{fig2}d) the recovery of the ferromagnetic order takes place in localised areas of the system due to the short-range atomic exchange interactions. This process leads to the nucleation of unstable and localised spin textures, namely \textit{magnon drops} (MD) \cite{turgut2016stoner}, following a non-equilibrium path in the spin configuration space. After a time-scale of hundreds of femtoseconds, during the so-called \textit{magnon coalescence} process (Region IV, green in Fig.\ref{fig2}b, and Fig.\ref{fig2}e) MD can scatter, split or merge until some equilibrium configuration is achieved\cite{maiden2014attraction}. Relaxing the spin system to RT by a heat-sink coupling (Region V in Fig.\ref{fig2}b), which occurs at the ns timescale, little change in the spin dynamics is observed and the skyrmion structures remain stable versus thermal fluctuations  at RT (Fig.\ref{fig2}f).
	
It is worth to emphasise that the  magnetisation dynamics leading to the generation of skyrmion lattices takes place in a time scale of picoseconds, following a rapid temperature change in the timescale of the order of 10 ps and below, i.e. it is a highly thermodynamically non-equilibrium path. Instead, slow temperature variations as the one used in Fig.\ref{fig1}d and \ref{fig1}b with zero-field cooling from a temperature above $T_C$=754.29K (see Appendix A) in a time scale of few nanoseconds always lead to a complex stripe domain-like configuration (see more results in Appendix B). This behaviour unveils the crucial role of the ultrafast character of the laser-induced nucleation of skyrmion lattices.

\begin{figure}[htb]
    \includegraphics [trim=0 0 0 0,width=\columnwidth]{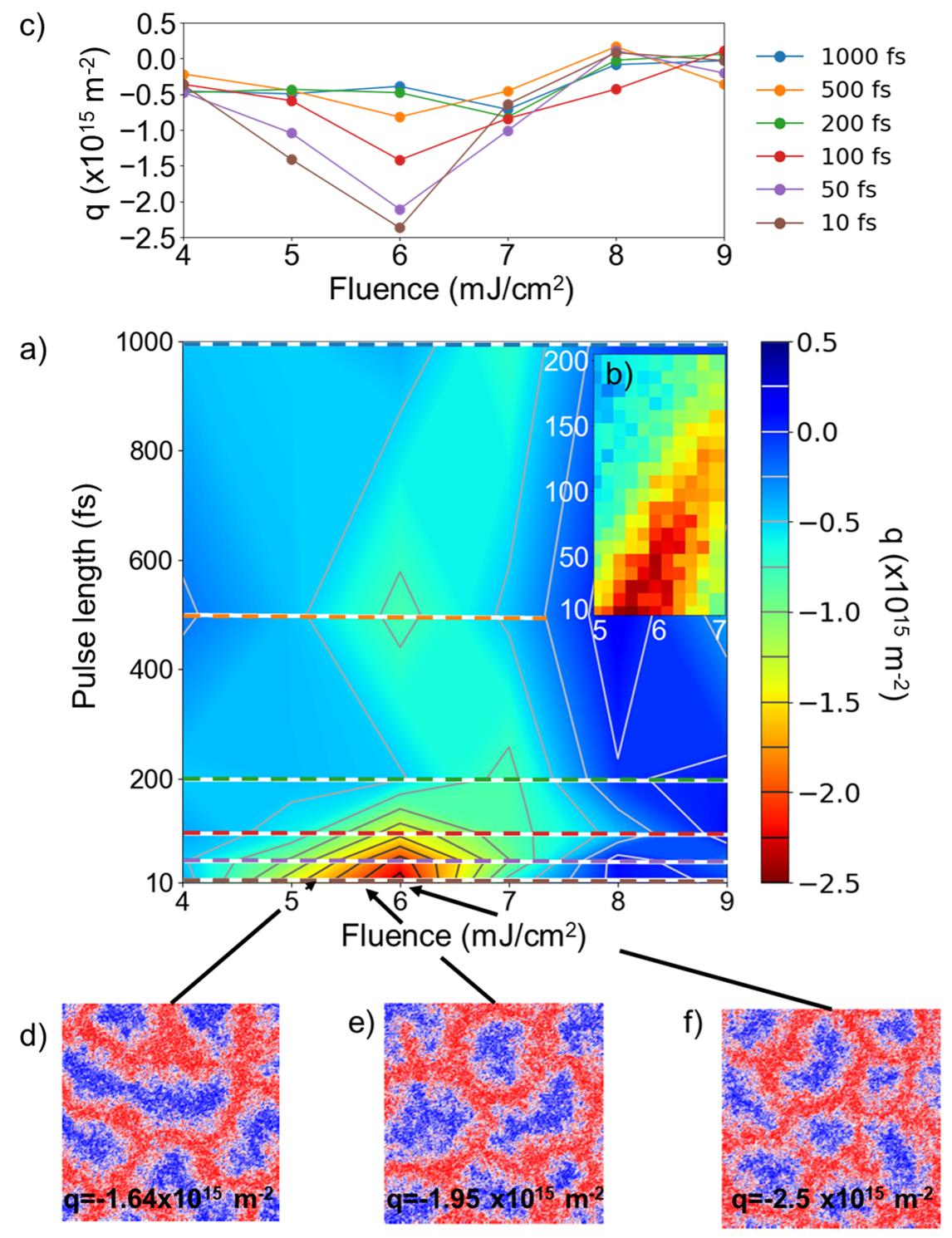}
    \caption{(Color online) {\bf Phase Diagram of skyrmion nucleation in zero external field}. a) Topological charge density $q$ (color scale) as a function of  laser fluence and pulse length. b) High resolution phase diagram showing the region with maximum topological density, averaged for 20 realisations; c) Topological charge density at selected pulse lengths as a function of fluence,   the data correspond to the horizontal lines in panel a); Panels d),e),f) show examples of spin configurations for different topological charge densities;}
    \label{fig3}
\end{figure}

\begin{figure*}[htb]
    \includegraphics [trim=0 0 0 0,width=\textwidth]{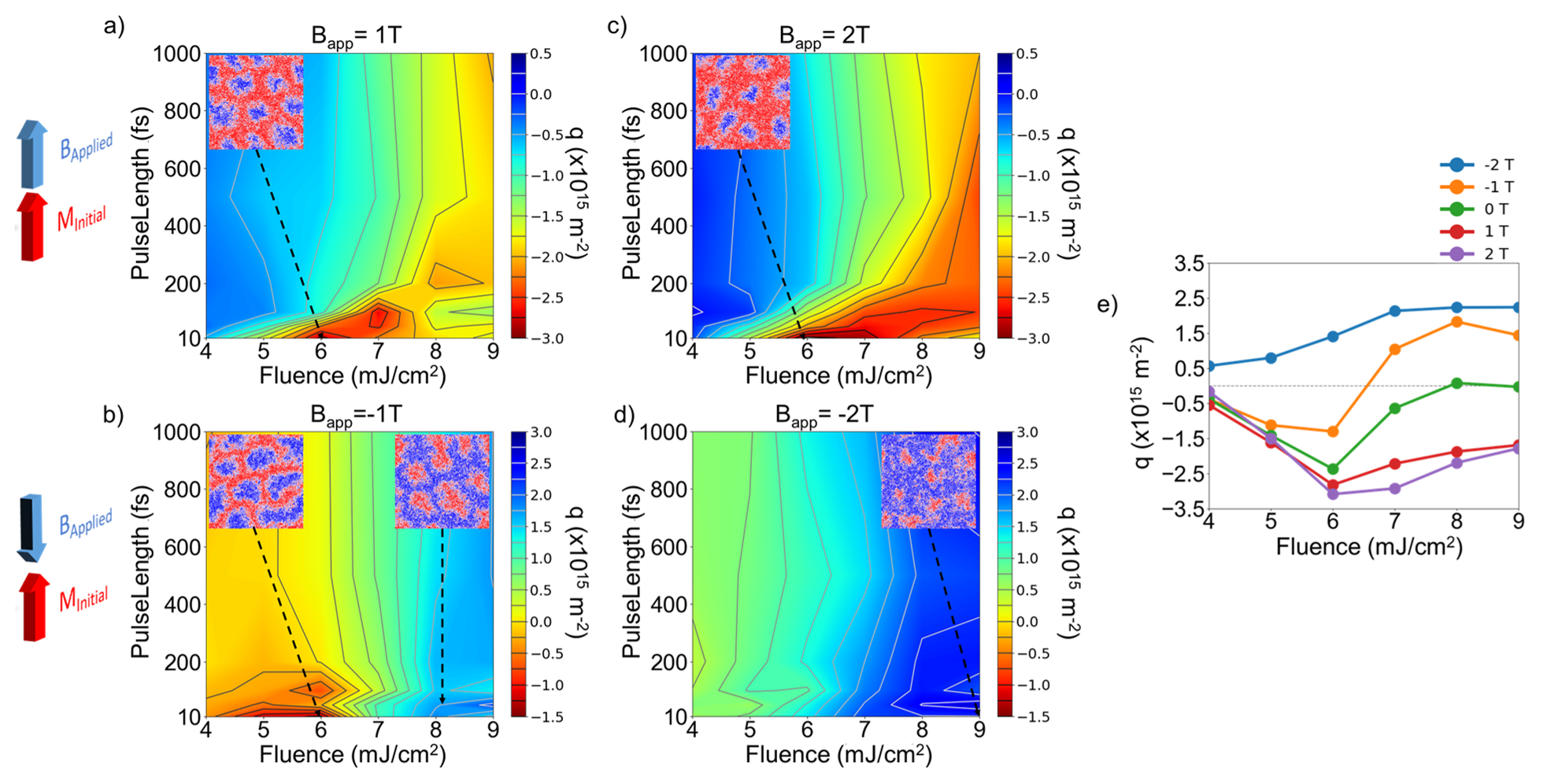}
    \caption{(Color online) {\bf Phase diagram when external fields are applied}. a) Phase Diagram when an external field of 1 Tesla is applied parallel to the initial magnetisation. b) Phase Diagram when an external field of 1 Tesla is applied antiparallel to the initial magnetisation. c)Phase Diagram when an external field of 2 Tesla is applied parallel to the initial magnetisation. d) Phase Diagram when an external field of 2 Tesla is applied antiparallel to the initial magnetisation. e)Topological charge density vs Laser fluence for selected external fields and for a fixed pulse width of 10 fs}
    \label{fig4}
\end{figure*}

To find the optimum conditions for the nucleation of skyrmion lattices, we have computed the phase diagram of the topological charge density $q$ as function of laser-pulse width $t_p$ and fluence $F_0$ in the absence (Fig.\ref{fig3}) and presence (Fig.\ref{fig4}) of a magnetic external field. This is computed  at a short time (250 ps) after the laser pulse is applied, but when all three subsystems(electrons, phonons and spins) are in mutual equilibrium, i.e. have the same temperature. To neglect the effect of a subsequent slow thermal relaxation, a vanishing heat-sink coupling (i.e. $\kappa_e$=0) is considered. Due to the stochastic nature of the  dynamics in the presence of temperature, the computed phase diagrams were averaged over 10 random realisations. It is important to note that the skyrmions can nucleate either with their core pointing parallel (P-Sk), i.e. $q>0$, or antiparallel (AP-Sk) i.e. $q<0$ to the initial magnetisation direction.

Fig.\ref{fig3} shows the  nucleation results in zero external field. The diagram  indicates that the skyrmion nucleation takes place for short pulse durations (below 100fs) and in a window of fluences around $F_0=6$mJ/cm\textsuperscript{2}. Fig.\ref{fig3}c shows  the topological charge density dependence on the laser fluence for selected laser pulse lengths (corresponding to the horizontal lines in Fig.\ref{fig3}a). It can be observed that $|q|$ increases initially with the fluence, showing a maximum  at a fluence of about 6 mJ/cm\textsuperscript{2}, which corresponds to 8 skyrmions generated in the simulated area (see Fig.\ref{fig3}f).  By analysing the spin configurations, a well defined skyrmion lattice can be obtained with a  minimum topological density of $|q|$ $\simeq$ 1.9 $\times$ 10\textsuperscript{15} m\textsuperscript{-2}. This value is considered here as a criterion for the nucleation of a skyrmion lattice. For smaller values of $|q|$, labyrinthic domains coexisting with skyrmions are obtained. A closer inspection of the magnetisation dynamics reveals that the successful nucleation of skyrmion lattices depends on two aspects: i) it is necessary to populate the sample with a sufficiently large number of MDs during the magnon localisation; ii) once the MDs are nucleated, they should gain topological protection so their merging is impeded. Thus, for short pulses and below $F_0\leq5$ mJ/cm\textsuperscript{2} the system is not demagnetised sufficiently to nucleate a large number of MD. At the same time, the topological protection is achieved with ultrafast cooling.
For fluences larger than $F_0>7$mJ/cm\textsuperscript{2} the energy deposited in the system is too high and the temperature fluctuations are too strong to allow stable MD and they merge into larger labyrinth-like domains. Further analysis and dicussions are present in appendices. For pulses longer than 100fs, the system stays longer at high temperature and thermal activation is therefore more likely to lead the system to reach the ground state. Thus, the competition of different factors defines a unique window for laser parameters.

In the following we analyse the behaviour of the skyrmion lattice formation as a function of external magnetic fields applied in different orientations. Fig.\ref{fig4} shows the nucleation phase diagrams under several magnetic fields that are applied parallel ($B_{app}>0$) or anti-parallel ($B_{app}<0$) to the initial magnetisation direction. It is worth noting that in the presence of external fields of these magnitudes, skyrmion lattices are stable ground states of the system (See Appendix B). Consequently, under long  heating and subsequent cooling (for pulses longer than $t_p \geq 200$ fs and high fluences), the system reaches the skyrmion state. In this case either the intermediate paramagnetic state is achieved or the skyrmion nucleation process goes through thermal activation from the ferromagnetic state over the energy barrier in which case the deposited laser energy and its duration should be sufficient to achieve high reversal probability in order to reach a skyrmion state.

Focusing on shorter pulses (below $t_p<200$ fs) for applied fields parallel to the initial magnetisation direction (Fig.\ref{fig4}a and \ref{fig4}c) we observe that the region where skyrmion lattices are obtained is substantially increased for fluences $F_0>F_C=6$mJ/cm\textsuperscript{2}. This behaviour can be ascribed to the hindering of the growth of domains oriented antiparallel to the field. At the same time, once the skyrmion is created, an external field parallel to their core increases the energy barrier separating it from the other states \cite{tejo2018distinct} and  the structure is more stable against fluctuations. 
The average topological charge densities  are larger than those obtained in the zero-field case. This feature is in the first place related to the smaller skyrmion sizes due to the influence of external fields, allowing a large number of structures in the sample (see inset in Fig.\ref{fig4}c). The effect of the applied field on the skyrmion size is investigated in Appendix D. Importantly, there is a substantial influence of the non-equilibrium excitation.  For example, in Fig.\ref{fig4}a the region with fluences larger than $F_0> 8$ mJ/cm\textsuperscript{2} and  for $t_p<200$fs does not show the formation of the stable skyrmion lattice, while for less intense pulses  skyrmion formation is obtained. This effect reveals once again that for pulses below $t_p<200$fs the system follows a non-equilibrium excitation and the magnon localisation and coalescence processes take place.

Figs.\ref{fig4}b and \ref{fig4}d show the nucleation phase diagrams for anti-parallel fields. Under an applied negative field, a ferromagnetic system can undergo ultrafast switching. In these cases, MD are nucleated with their core pointing parallel to the initial magnetisation. Thus, P-Sk lattices arise and the topological charge density carries a positive sign. This  can be observed in Fig.\ref{fig4}b and \ref{fig4}e (yellow line), where both positive and negative polarisations of skyrmion core can be obtained with field $B_{app}=-1$T for short pulse lengths. Negative $q<0$ values are consistent with the fact that larger fluences  favor the magnetisation switching. For $B_{app}=-2$T, the Co trilayer magnetisation  always switches its orientation. The  effect of external fields is summarised in Fig.\ref{fig4}e for a fixed ultrashort pulse width of $t_p=50$ fs.

In conclusion we have modelled the linearly-polarised laser- (heat-)induced nucleation of skyrmion lattices in epitaxial cobalt-based asymmetric trilayers with perpendicular magnetic anisotropy. The system presents a ground state with labyrinth domains. The computed  phase diagram for topological charge showed a nucleation window in terms of laser pulse intensity and duration, corresponding to ultra-fast lasers pulses shorter than $t_p<200$fs. This shows the importance of highly thermodynamically non-equilibrium path for skyrmion generation. This path involves magnon drop nucleation after the sample demagnetisation, and their topological protecion during the magnon coalescence. Additionally, external fields enlarge the nucleation window, reduce skyrmion sizes and change the nucleation probability. Thus, we propose the important techniques and protocols to influence the desired skyrmion densities by a combined action of the laser pulse and external fields. Our results are important for the control of skyrmion nucleation for novel applications.

\begin{acknowledgements}
The authors acknowledge the networking opportunities provided by the European COST Action CA17123 "Magnetofon" and in particular the short-time scientific missions awarded to P.O.-R. and M.S. The work in Spain was supported by the Regional Government of Madrid through Project P2018/NMT-4321 (NANOMAGCOST-CM) and by the Spanish Ministry of Economy and Competitiveness (MINECO) through Projects SKYTRON (FIS2016-78591-C3-1-R and FIS2016-78591-C3-3-R) and FUNSOC (RTI2018-097895-B-C42). This project has received funding from the European Union’s Horizon 2020 research and innovation programme under grant agreement No 737093 FEMTOTERABYTE and under the FLAG-ERA JTC2019 Project SOgraphMEM (MINECO PCI2019-111867-2). IMDEA Nanoscience Institute is supported by the “Severo Ochoa” Programme for Centres of Excellence in R\&D, (MINECO grant SEV-2016-0686).

\end{acknowledgements}
\appendix
\section{Curie Temperature.}
\label{appA}
\begin{figure}[htb]
    \includegraphics [trim=0 0 0 0,width=\columnwidth]{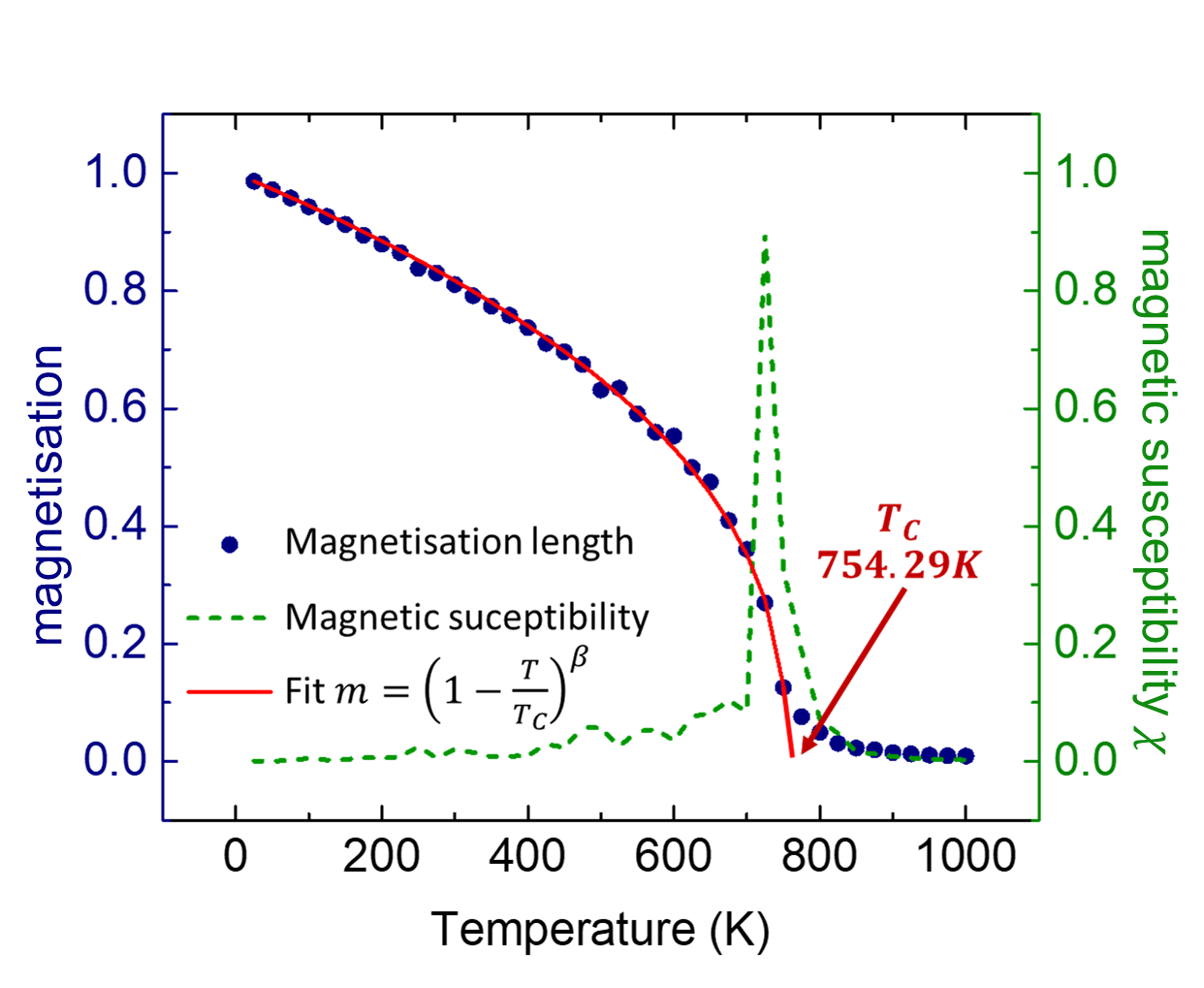}
    \caption{(Color online) \textbf{Curie Temperature.} Simulated magnetisation $m=M/M_s$ (blue dots, left axis) and magnetic susceptibility (green dashed line, right axis) versus temperature.  The Curie Temperature of the system is obtained by fitting the simulated $m$ vs $T$ curve  to the expression $m(T)=(1-T/T_C)^\beta$ (red solid line).
    }
    \label{fig1app}
\end{figure}

We obtained the Curie Temperature of the system $T_C$ by reproducing the reduced magnetisation $m=M/M_S$  dependence with the temperature $T$. The Dzyaloshinskii-Moriya Interaction (DMI) (having negligible influence on $T_C$) is deactivated in this case as it produces the nucleation of labyrinth domains with no net magnetisation. The simulation starts from an initial saturated state pointing along the positive out-of-plane direction  (i.e. $m_Z = +1$) increasing the temperature  in steps of $\Delta T = 25$ K and equilibrating the system at each temperature by solving the stochastic LLG equation with a time step  $\Delta t=10^{-16}$ s during  100ps. After the initial equilibration stage, the magnetisation is averaged over 100ps. In figure \ref{fig1app}, the averaged magnetisation $m$ is plotted versus temperature (blue dots) $T$. The Curie temperature $T_C$ is obtained by fitting (red solid line) the resulting  curve to the following expression.
\begin{equation}
\label{eq3}
    m(T)=(1 - \frac{T}{T_C})^{\beta}
\end{equation}
giving $T_{C} = 754.29$ K and $\beta = 0.39$ for the considered 3 monolayers of hcp-Cobalt system. The susceptibility of the system is calculated as following:
\begin{align}
    \chi_{\alpha}= &\frac{\mu_i}{k_B T} (<m_{\alpha}^2> - <m_{\alpha}>^2),~\\ &m_{\alpha} = {m_x,m_y,m_z,|m|}
\end{align}
The divergence point of susceptibility associated to the ferromagnetic to paramagnetic phase transition agrees with the fitted Curie Temperature.

\section{Comparison of magnetic states after slow field-cooling and laser-induced processes.}
\label{appB}
\begin{figure}[htb]
    \includegraphics [trim=0 0 0 0,width=\columnwidth]{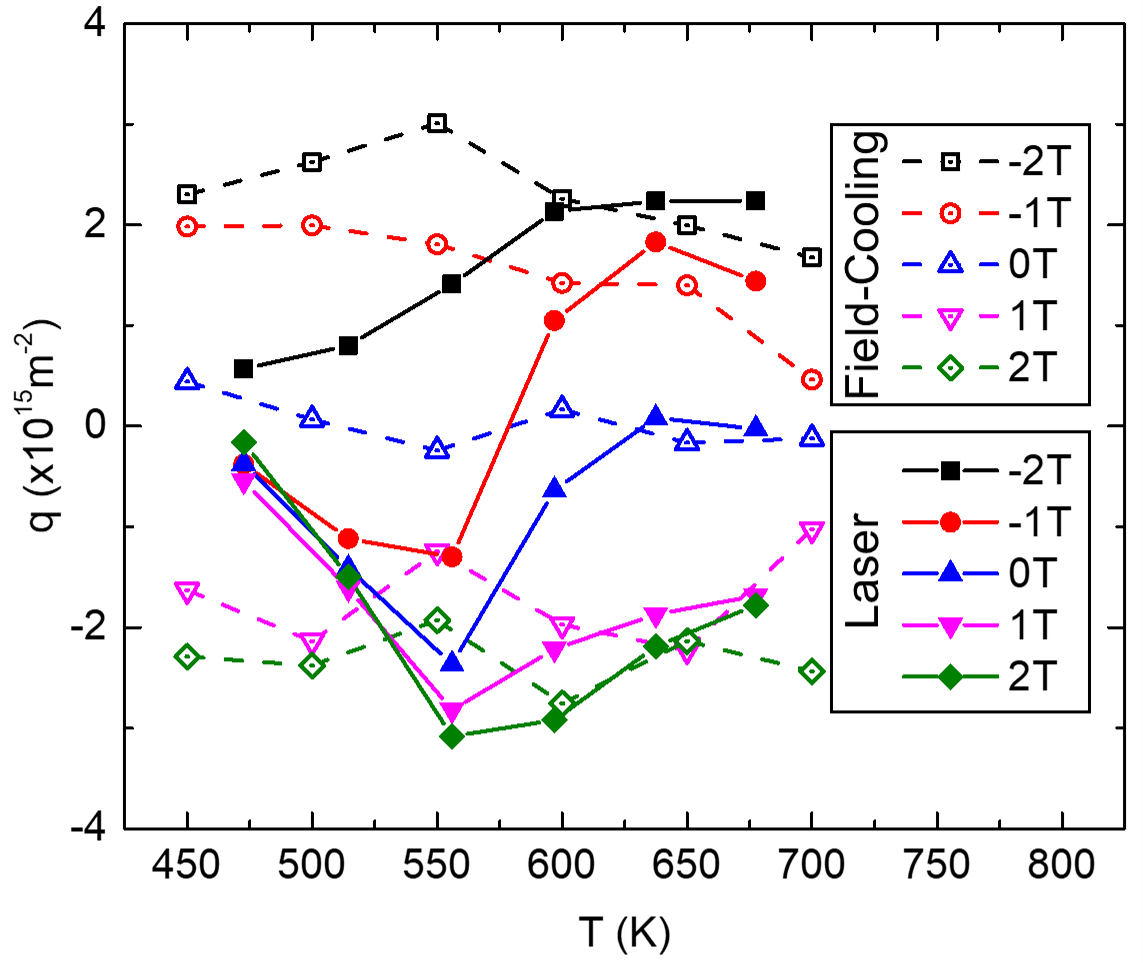}
    \caption{(Color online) \textbf{Laser-induced vs Field-Cooling processes.} Comparison of the topological charge density $q$ obtained from ultrafast (filled symbols) and quasi-equilibrium field-cooled (empty symbols) processes as a function of the final system temperature. The simulations were carried out under different external fields. The lines are guidelines.}
    \label{fig2app}
\end{figure}

\begin{figure}[htb]
    \includegraphics [trim=0 0 0 0,width=\columnwidth]{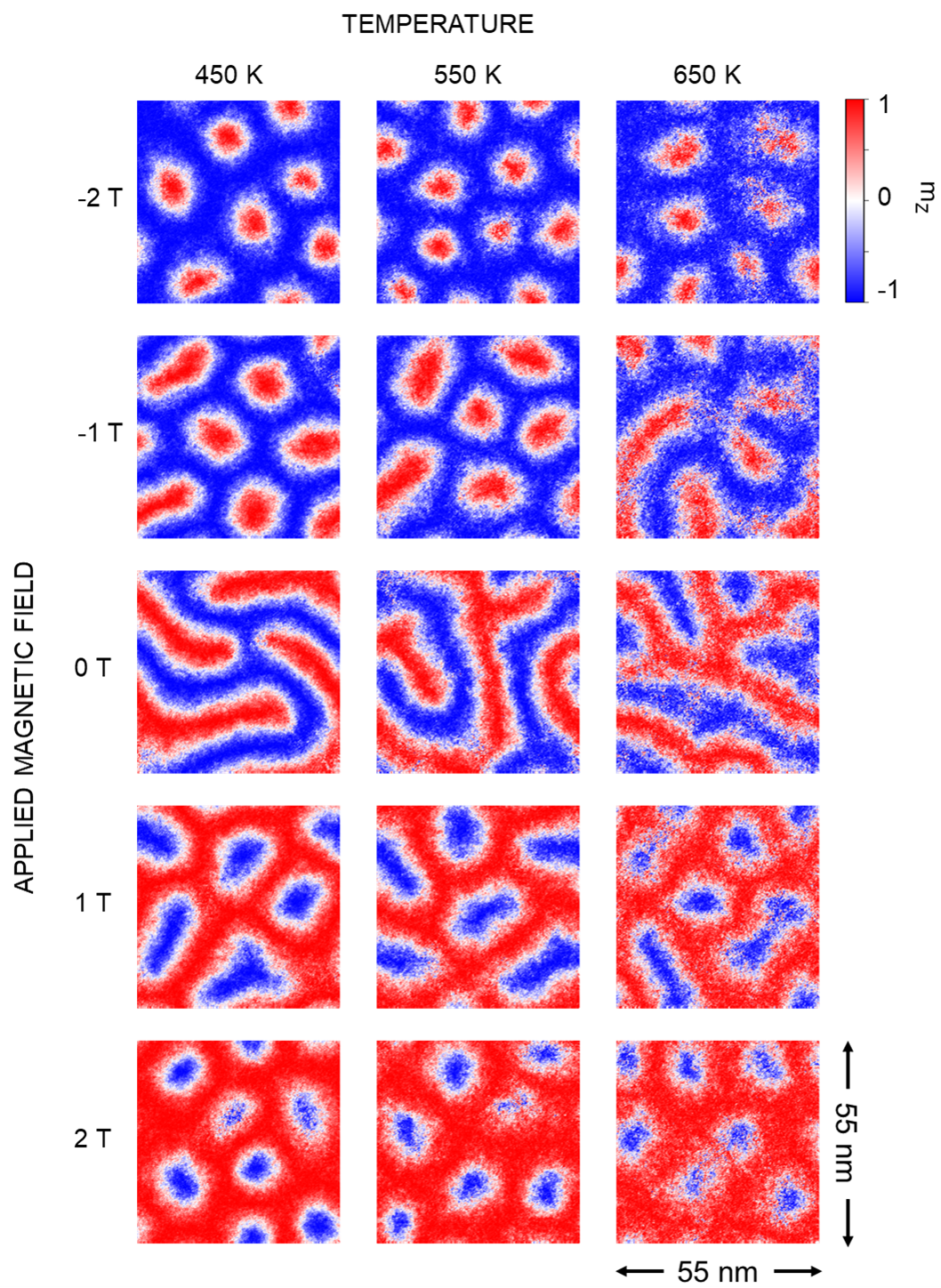}
    \caption{(Color online) \textbf{Ground states.} Obtained after slow field-cooling processes at different applied fields.}
    \label{fig3app}
\end{figure}

The role of the non-equilibrium character of the laser-induced nucleation of skyrmion lattices is emphasized by contrasting them with the ground states of the system when following a slow field cooling process (FCP) from an initial paramagnetic state. In these simulations the system is initially placed at a temperature $T$=1000K  and is cooled down at a constant rate of $dT/dt=175$K/ns until different final temperatures are reached. In all cases the LLG equation is integrated during a total simulated time of 4.5ns. The final temperatures were selected to be comparable  to the final quasi-equilibrium temperatures  after a laser pulse,  when both the electronic and the lattice temperatures have reached thermal equilibrium. It is important to remark that the latter temperature  depends only on the laser fluence $F_0$ and not on the laser pulse width. The ground states were obtained under different external fields. In order to minimise the thermal noise and highlight the spin textures, the final states  have been averaged during the last 100ps of simulation. In figure \ref{fig2app} we present the comparison of the topological charge density computed when following a FCP  and an ultrafast laser heating. Fig.\ref{fig3app} presents the obtained configurations following the FCP while the configurations resulting from ultrafast heating are presented in the main text, Figs.\ref{fig3} and \ref{fig4}. 
%It is important to mention that while the results related to the laser pulse have been averaged using 10 different sets of simulations, the FCP simulations have been calculated just once.

For the FCP processes the topological charge density of the system increases with the strength of the external applied field. This result is expected given the fact that under the influence of the external fields  smaller skyrmions nucleate. Note that the behaviour is symmetric with respect to the sign of the applied field. In the absence of external fields, the topological charge vanishes and takes values close to zero indicating that no skyrmion lattice can be nucleated by following quasi-equilibrium paths. 

When compared to the results obtained after applying a laser pulse, the first remarkable difference that can be seen is for temperatures lower than $T<$550K (Fluences  $F_0<6$ mJ/cm\textsuperscript{2}). In that range the topological charge densities obtained by means of  the laser pulse heating are smaller than those obtained after a field-cooling process. This is due to the fact that these laser fluences are not high enough to increase the temperature of the system above or even close to $T_C$. Thus, the system retains the ferromagnetic state and only a small amount of magnon drops are  nucleated in the film. At the same time in the simulated FCP, the system always starts from a paramagnetic state and a large number of magnetic domains are nucleated in all cases.  

Secondly, when final temperatures are close to $T\approx 550$K  ($F_0=6$mJ/cm\textsuperscript{2}), the skyrmion lattice is nucleated with ultrafast heating for the cases of no field and positive external fields while for FCP the effect of the field is necessary in order to obtain a skyrmion lattice. The main difference introduced with the laser pulse is the large amount of magnon drops nucleated during the far from equilibrium magnetisation process. These magnon drops are able to survive as the rapid decrease of the temperature impedes their destabilisation and allow their growth.
Note that the skyrmion densities obtained in the absence of external fields (blue solid triangle at $T\approx550$K) are close to those obtained when a positive field  is applied (pink solid triangle and green solid diamond at $T\approx550$K). 
This may be attributed to the nucleation of a large number of magnon drops which achieve topological protection before they meet. Thus, when external positive fields are applied it is possible to prevent some merging events during the magnon coalescence stage. 
The above is in contrast to the results of the FCP which show a larger difference in topological charge under an applied field.

 If we apply negative external fields (i.e. pointing antiparallel to the initial magnetisation) two different scenarios are observed. For  $B_{app}=-1$T (red solid circle at $T\approx550$K) we observe a decrease in the skyrmion density $q$. In this case, the external field is promoting the growth of the magnon drops and their merging during the coalescence stage. As a result a smaller amount of large skyrmions survive (see inset in Fig \ref{fig4}b)). When  $B_{app}=-2$T (black solid square at $T\approx550$K) the skyrmion density changes its sign meaning that now the skyrmions have their core pointing opposite to the previous cases. This may happen because of two reasons. First, it might happen that due to the strong external field, the sample exhibits a magnetisation reversal process. In that case, the magnon drops would be nucleated with their core pointing parallel to the initial magnetisation and we would find the opposite situation as in $B_{app}=+2$T with approximately the same absolute value of $|q|$. However, we observe that $|q|(B_{app}=+2)>>|q|(B_{app}=-2)$ and thus the above explanation should be discarded. A second scenario can be described taking into account that the external field promotes the growth and merging of magnon drops during the magnon coalescence. In such case, small regions of the system with magnetisation pointing parallel to the initial magnetisation might survive, playing the role of parallel skyrmions. This scenario agrees with the predicted difference of the absolute value of the topological charge compared with its counterpart for $B_{app}=2$T case. 

Finally, if we increase the laser fluence ($F_0>6$mJ/m\textsuperscript{2}) and  reach temperatures higher than $T=600$K, we find that the results of the ultrafast and equilibrium excitations are very close. We argue that in these cases the final temperature is so large that the recovery of magnetic correlations needs long time, thus, the magnon localisation stage never takes place.

%\appendix
\section{Influence of magnon localisation and coalescence processes.}
\label{appC}
\begin{figure}[htb]
    \includegraphics [trim=0 0 0 0,width=\columnwidth]{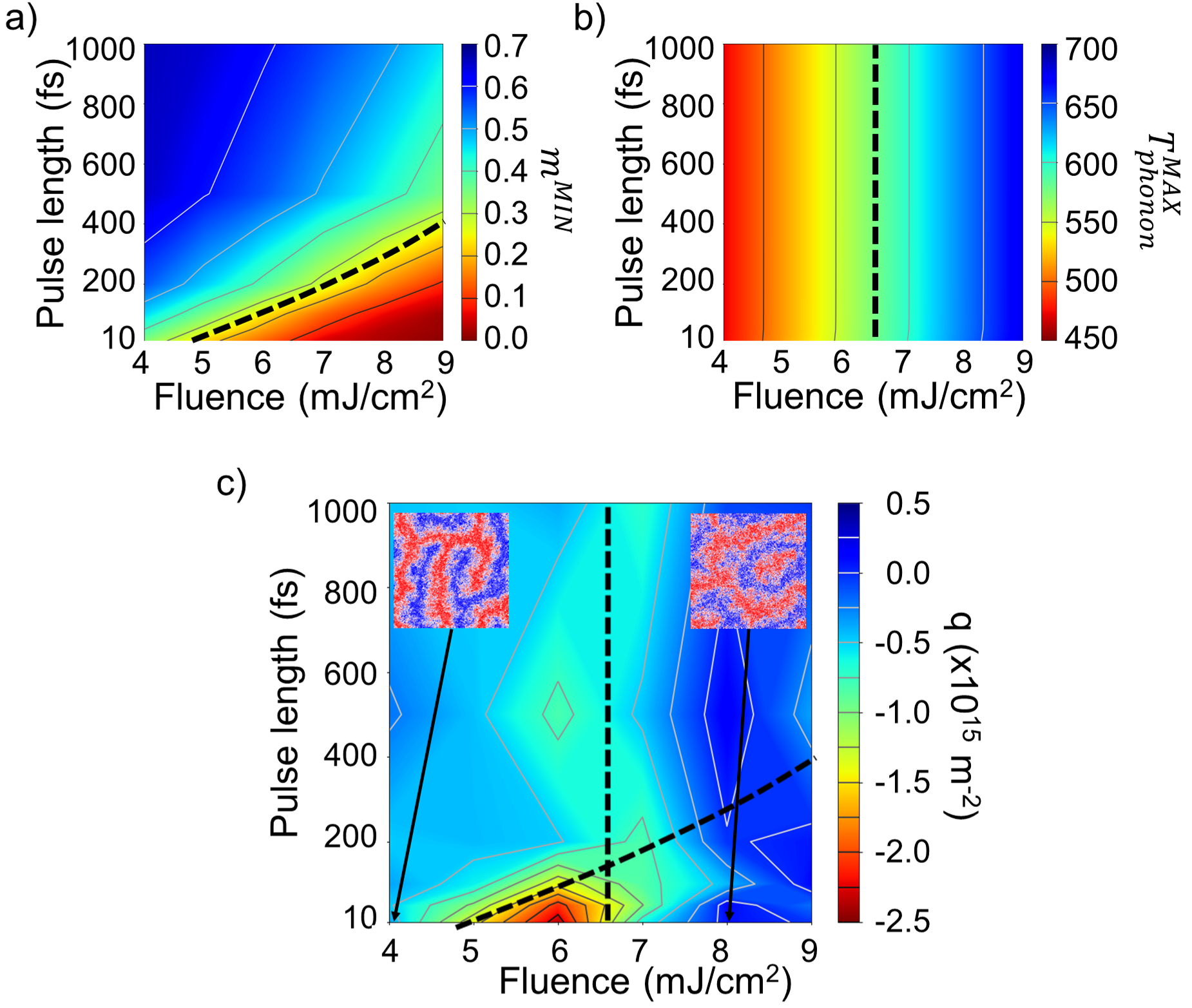}
    \caption{(Color online) {\bf Relation of the induced demagnetised state and final equilibrium temperature with the computed skyrmion densities}.  a) Magnetisation of the system after the laser pulse
   application. The dashed black line represents a constant minimum magnetisation of $m^{MIN} = 0.25$. b) Final equilibrated phonon-electron temperature. The dashed black line represents a constant equilibrium temperature of $T_{phonon}^{MAX}=580K$. c) Topological charge density. The dashed black lines are transferred from panels a) and b) and limit the region of interest. The insets show particular cases in which large labyrinthic-like domains are promoted.}
    \label{fig4app}
\end{figure}
The influence of the magnon localisation process and the quasi-equilibrium phonon-electron temperature on the nucleation of skyrmion lattices can be seen in Fig. \ref{fig4app}. In panel \ref{fig4app}a we show the minimum magnetisation reached by the system after the laser pulse is applied. In panel \ref{fig4app}b we present the final quasi-equilibrium phonon-electron temperature due exclusively to the effect of the laser. The far from equilibrium nature of the magnon localisation process requires a large level of demagnetisation in order to nucleate a larger number of magnon drops. On the other hand and during the magnon coalescence phase, the thermal fluctuations affects the achievement of topological protection, and the final equilibrium temperature (that matches the maximum temperature reached by the phononic subsytem) should not overcome a certain value to allow the protection of the domains. In panels a) and b) the dashed black lines represent a constant minimum magnetisation of $m^{MIN} = 0.25$ and a constant equilibrium temperature of $T^{MAX}_{phonon} = 580 K$ respectively. If we transfer these lines to the computed phase diagram of the topological charge density (Fig.\ref{fig4app}c) it can be observed that the region of interest in which the skyrmion lattices are nucleated is well delimited within the lines. Thus, it can be interpreted that for fluences  $F_0>6.5$mJ/cm\textsuperscript{2}, the high temperatures reached by the system prevents the topological protection of the magnetic domains and they merge into larger and labyrinthic-like domains (see right inset in Fig.\ref{fig4app}c). For fluences $F_0<6.5$mJ/cm\textsuperscript{2} if the minimum magnetisation is above $m^{MIN} > 0.25$ we are not demagnetising the system enough to promote the nucleation of a sufficient number of magnon drops and only a few of them arises and grow once again as large and labyrinthic-like domains(see left inset in Fig.\ref{fig4app}c).

\section{Skyrmion radii.}
\label{appD}
\begin{figure}[htb]
    \includegraphics [trim=0 0 0 0,width=\columnwidth]{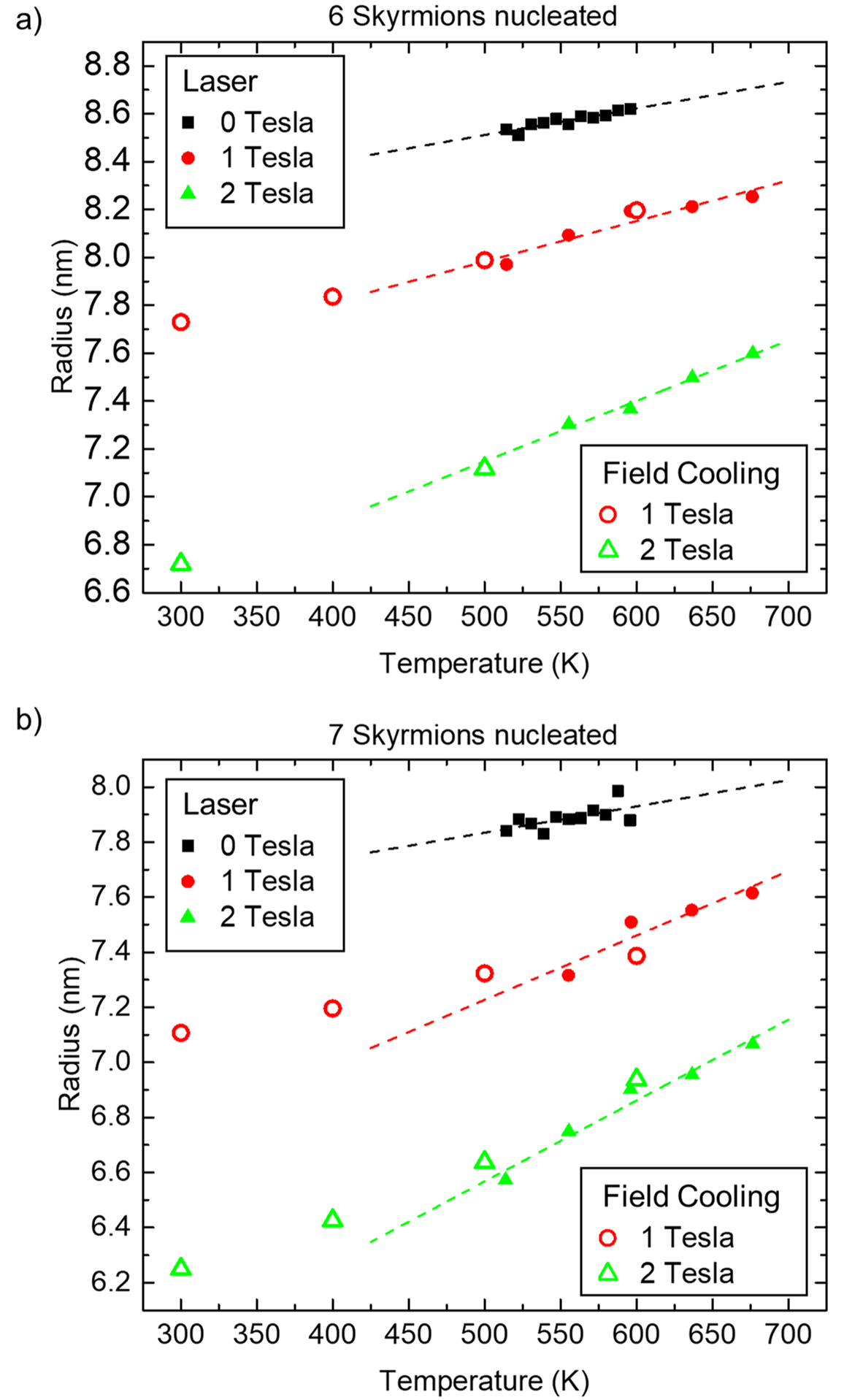}
    \caption{(Color online) {\bf Skyrmion radius vs temperature.} The skyrmion radii $R_{sk}$ are computed considering the relation $R_sk=\sqrt{(m+\frac{q}{|q|})/{2q\pi}}$ for the cases where a well defined skyrmion lattice is obtained (i.e. $q>1.9 \times 10^{15}$m\textsuperscript{-2}). The dashed lines are linear fits added for illustrative purposes.}
    \label{fig5app}
\end{figure}

The skyrmion radii can be computed by assuming that in the cases where well defined skyrmion lattices are nucleated (i.e. $q\geq1.9 \times 10^{15}$m\textsuperscript{-2}), these adopt an almost circular shape and the domain wall widths are negligible compared to the skyrmion sizes. Taking this into account, the reduced magnetisation is related to the area of the sample with magnetisation  pointing upwards(downwards) $A^{\uparrow}$($A^{\downarrow}$) by $m=\frac{A^{\uparrow}-A^{\downarrow}}{A^{\uparrow}+A^{\downarrow}}=1-\frac{2A^{\downarrow}}{A}$; where $A=A^{\uparrow}+A^{\downarrow}$ is the total area of the sample. In a sample populated with skyrmions whose core is pointing downwards i.e. the negative values of the OOP direction $A^{\downarrow}=N_{sk}\pi R_{sk}^{2} = An_{sk}\pi R_{sk}^{2}$. The magnetisation is thus related to the skyrmion density as $m=1-2n_{sk}\pi R_{sk}^{2}$. Considering the polarization of the core of the skyrmions, the latter relation can be generalized as $m=-P_{core}+2n_{sk}P_{core}\pi R_{sk}^{2}=-\frac{q}{|q|}+2q\pi R_{sk}^{2}$. This allows to approximately evaluate the average skyrmion size from magnetisation.

Our first observation is related to the fact that the skyrmion radius depends on the number of nucleated skyrmion due to the finite size of the modelled system and elevated non-equilibrium temperatures. This may happen also in the experimental situation due to the finite size of the laser spot. Secondly, while during the laser-induced nucleation we got from 5 to 10 nucleated skyrmions, in the adiabatic heating/cooling simulations we get a smaller dispersion in the number of nucleated skyrmions (from 6 to 8).

In figure \ref{fig5app} we show the  dependence of the computed skyrmion radii for the selected cases in which 6 skyrmions(\ref{fig5app}a) and 7 skyrmions (\ref{fig5app}b) are nucleated within the simulated sample (related to skyrmion densities of $q=1.98\times10^{15}$m\textsuperscript{-2} and $q=2.31\times10^{15}$m\textsuperscript{-2} respectively). The radii were computed for the cases in which both, the laser-induced nucleation (solid symbols) and the adiabatic field-cooling processes (empty symbols) were followed and they have been averaged over various sets of simulations.  With illustrative purposes, linear fits of the light-induced nucleated skyrmion radii have been added(dashed lines). 

As expected, we can observe that the skyrmion radius decreases as the external applied field is increased. Note the background switching for $2T$ applied field. It also increases with temperature following the predictions of Ref.\cite{Tomasello2018}.
An interesting result reveals that in the high-temperature region the skyrmion radius follows a linear trend with temperature. 

Interestingly, the slope of the in-field dependence of the skyrmion radius with temperature is higher than at zero. Thus, when an external field $|B_{app}|=2$T is applied (green symbols) the slope is increased compared to the $|B_{app}|=1$T (red symbols) and $|B_{app}|=0$T (black symbols) cases. This is in agreement with the predictions of Ref.\cite{Tomasello2018} that the stable skyrmion have a more pronounced temperature dependence that the metastable ones.

Comparing panel a (6 nucleated skyrmions) and panel b (7 nucleated skyrmions) we also observe that the skyrmion radius decreases as a function of the skyrmion density.  This result follows from the fact that skyrmions are constrained due to finite size of the simulated system and their mutual interaction. In the absence of periodic boundary conditions, an expansion of the magnetic skyrmions is expected with time, along with a convergence of the radii towards a single value that depends only on the magnetic parameters of the sample. When comparing the results obtained following the laser-induced protocol (solid symbols) and following the field-cooling processes (empty symbols) we observe that in the case where 6 skyrmions are nucleated following a field-cooling process match perfectly with the trend shown by the radii of the light-induced nucleated skyrmions. However, in the case in which 7 skyrmions are nucleated, the trends of the radii in both, the light-induced and the field-cooling induced nucleated skyrmions do not match. This mismatch might arise due to the constraint described above which may affect the incidence of the high temperatures in the fluctuations of the topological protection. Thus, in a more stressed (constrained) skyrmion, the large temperatures induces larger fluctuations in $q$ affecting our computed radii. If this is the case, and given the good agreement of the linear fit exposed in the case where 6 skyrmions are nucleated, we can assume that $q=1.98\times10^{15}$m\textsuperscript{-2} is the equilibrium value of the skyrmion density hosted by a system with the defined magnetic parameters. To conclude this analysis, we would like to point out that in our model the magnetostatic interaction is not considered. While we do not expect this interaction to play a significant role in the nucleation, as it is fully screened during the magnon localisation process and even it would help to prevent the merging of topologically protected skyrmions during the magnon coalescence phase, is it expected that the equilibrium skyrmion sizes will be increased.

%\begin{acknowledgments}
%\end{acknowledgments}
\bibliography{references.bib}

\end{document}